\documentclass[12pt,a4paper]{article}
\usepackage{amsmath}
\usepackage{amssymb}
\usepackage{amsxtra}
\begin{document}
\newtheorem{theorem}{Theorem}
\newtheorem{axiom}{Axiom}
\newtheorem{corollary}{Corollary}

\title{A correspondence between standard model fermions and degrees of
freedom of polycrystalline materials} 
\author{Ilja Schmelzer}
\maketitle
\sloppy \sloppypar
\begin{abstract}

We identify natural degrees of freedom of polycrystalline materials --
affine transformations of grains -- with those of a three-dimensional
lattice theory for $(T\otimes\Omega)(\mathbb{R}^3)$.  We define a lattice
Dirac operator on this space and identify its continuous limit with the
free field limit of the whole fermionic sector of the standard model.
Fermion doubling is used here as a tool to obtain the necessary number
of steps of freedom.  The correspondence extends to important
structural properties (families, colors, flavor pairs,
electromagnetic charge).  We find a lattice version of chiral symmetry
similar to the Ginsparg-Wilson approach.

This correspondence suggests to propose a ``polycrystalline ether''.
Combined with GLET, a general Lorentz ether theory of gravity with GR
limit, this becomes a concept for a theory of everything.  The
extension to gauge fields is the major open problem and requires new
concepts.

\end{abstract}

\tableofcontents

\section{Introduction}

The main idea of this paper may be formulated like the prototype of
crank physics: As a proposal of a {\em the theory of everything\/}
which may described in a single sentence: {\em We have absolute time
and an Euclidean space, which is filled with a polycrystalline
ether.\/} To add further evidence in this direction, the author almost
fits into the ``engineer'' pattern typical for cranks: He has no PhD
and works in the domain of mesh generation for scientific computing.

For those who nonetheless have not yet stopped reading, let's start.
The key of this paper is a lattice version of the Dirac operator on
$(T\otimes\Omega)(\mathbb{R}^3)$.  Here $\Omega(M^n)$ denotes the exterior
bundle (de Rham complex) of a manifold $M^n$.  For a given metric
$g_{\mu\nu}$ on $M^n$ Hodge theory (see app. \ref{Hodge}) defines not
only a Dirichlet operator $\Delta$ on $\Omega(M^n)$, but also a
natural square root $D$, $D^2=\Delta$ called ``Dirac operator''.
There exists also a square root for the operator $\square =
\partial_t^2-\Delta$ which has the form $\hat{D}=\gamma^0\partial_t -
D$. We define also a complex structure on $(T\otimes\Omega)(\mathbb{R}^3)$
which corresponds to a quaternionic structure on
$\Omega(\mathbb{R}^3)$.

We consider a simple lattice discretization of this three-dimensional
Hodge theory Dirac operator for Euclidean metric, leaving time
continuous.  The key observation is the identification of the large
distance limit of this lattice equation with the free field limit of
the whole fermionic sector of the standard model.  The effect which
allows to reach this identification is known as ``fermion doubling''
in lattice theory.  If we try a naive (central differences)
discretization of the Dirac equation and then consider the large
distance limit we observe that we obtain not only the fermion which we
have tried to put on the lattice, but also highly oscillating
solutions which fulfill the same equation -- so called doublers or
spurious modes.  This doubling effect happens in each direction.  Once
we discretize only in spatial direction, we obtain a factor $2^3=8$.
Now, the space $(T\otimes\Omega)(\mathbb{R}^3)$ has dimension 24.  Together
with the doubling factor 8 this gives exactly what we need to model
the 24 standard model fermions, with their 8 real (4 complex) degrees
of freedom for each fermion.

The identification does not reduce to $192 = 192$, but gives all the
structural properties of the standard model we would like to have: A
representation of the standard Dirac matrices $\gamma^\mu$ which
allows to identify fermions, three fermion families, flavor pairs,
three colors for quarks.  We have also a natural candidate for
electromagnetic charge in terms of the number of oscillating
directions.  Moreover, we have found a version of chiral symmetry on
the lattice so that the lattice Dirac operator has chiral symmetry.
It is defined by a pair of lattice operators $\gamma^5$,
$\tilde{\gamma}^5$ so that chiral symmetry for an operator $O$ means
${\gamma^5}O+O\tilde{\gamma}^5=0$, ${\tilde\gamma^5}O+O{\gamma}^5=0$.
Such a notion of chiral symmetry on the lattice is similar (but not
identical) to the generalization of chiral symmetry used in the
Ginsparg-Wilson approach to chiral gauge theory.

On the other hand, we can identify on the kinematical level our
discretization of $(T\otimes\Omega)(\mathbb{R}^3)$ with the degrees of
freedom of a classical polycrystalline material.  All what we need for
this identification is the granular structure, which defines a spatial
lattice, and the idea that the deformation of the granular structure
may be described by an affine transformation\footnote{Here and in the
following, if not mentioned otherwise, Latin indices $i,j,k,\ldots$
vary over $x,y,z$, Greek indices $\kappa,\lambda,\mu\nu,\ldots$ over
$t,x,y,z$.}  $q^i_\mu(n)=(q^i_0(n),q^i_j(n)) \in A(3)$ for each grain
$n$.  Together with the related momentum variables
$p^i_\mu(n)=(p^i_0(n),p^i_j(n)) \in A(3)$ we obtain the same 24
degrees of freedom per lattice node, and, again, with preservation of
the most important structural properties:

\begin{equation}
\psi^i(n) = q_0^i(n)+p^i_j(n)dx^j+q^i_j(n)*dx^j+p^i_0(n)dxdydz
\end{equation}

(with $*dx^i=\frac{1}{2}\varepsilon_{ijk}dx^jdx^k$).  

The extension of this kinematical identification to dynamics faces an
old problem of ether theory.  The Dirac equation is almost equivalent
to the d'Alembert equation, which is already a classical wave
equation.  But in classical condensed matter we usually have different
speeds for longitudinal and transversal waves, while this d'Alembert
equation gives the same speed for all waves.  We propose an ``atomic
ether'' variant which allows to motivate the d'Alembert equation as a
free field limit.  But this variant has less predictive power in
comparison with the ``polycrystalline ether'': The number of different
types of atoms -- four -- we have to put in by hand.

Then we consider the question of extension of the whole correspondence
to gauge fields.  Unfortunately, the standard Wilson approach to gauge
fields on the lattice does not work.  The reason is that
multiplication with $i$, which is assumed to be a pointwise operation
in this approach, does not have these properties in our case.  Real
and imaginary components of a fermion are located on different lattice
nodes.  Therefore the lattice versions of the operators $i$ as well as
$\gamma^5$ cannot be pointwise operators.  This may be a feature, not
a bug: It prevents the application of standard standard no-go theorems
like the famous Nielson-Ninomiya \cite{NN} theorem.  But this is an
open question for future research.

Nonetheless we can derive some general principles how gauge fields
have to be handled in an ether-theoretical approach.  These general
principles differ from those in the standard relativistic approach in
essential points.  The major difference is that we have no
factorization (BRST cohomology) but the gauge degrees of freedom have
to be physical degrees of freedom.  Thus, we have to go back before
the Gupta-Bleuer \cite{Gupta50},\cite{Bleuer} proposal to use an
indefinite Hilbert space structure in the ``big'' space to obtain
manifest Lorentz symmetry.

The ``polycrystalline ether'' proposal presented here nicely fits into
a general Lorentz ether theory (GLET) for gravity proposed by the
author in \cite{SchmelzerAnnals}.  We introduce this theory shortly in
app. \ref{GLET}.  In this theory, the gravitational field $g_{\mu\nu}$
is identified with density $\rho$, velocity $v^i$ and pressure tensor
$p^{ij}$ of an ether in a classical Newtonian framework.  The
identification is a variant of the ADM decomposition:

\begin{eqnarray*}
g^{00} \sqrt{-g} &= t^{00} =& \rho \\
g^{i0} \sqrt{-g} &= t^{i0} =& \rho v^i \\
g^{ij} \sqrt{-g} &= t^{ij} =& \rho v^i v^j + p^{ij}
\end{eqnarray*}

GLET describes only a general framework.  It allows, but does not
specify further ``inner steps of freedom'' -- fields $\varphi^m(x)$
which describe the material properties of the ether.  These fields
$\varphi^m(x)$ are identified with matter fields.  The Lagrangian
which follows from this general theory is

\begin{equation}
L = L_{GR}(g^{\mu\nu}) + L_{matter}(g^{\mu\nu},\varphi^m)
  - (8\pi G)^{-1}(\Upsilon g^{00}-\Xi (g^{11}+g^{22}+g^{33}))\sqrt{-g}
\end{equation}

The only restriction which follows for the matter Lagrangian
$L_{matter}$ is independence from the preferred coordinates -- the
Einstein equivalence principle.  Thus, GLET defines an ideal framework
for an ether-based theory of everything.  The final ether theory of
everything has to specify the material properties of the ether and to
derive from this the Lagrangian of the standard model or some
generalization.  Our ``polycrystalline ether'' nicely fits into this
scheme: The specification ``polycrystalline'' gives the standard model
fermionic particle content.

We consider some steps into the direction of unification of our flat
space Dirac operator with GLET: We remember shortly standard Hodge
theory which defines the continuous Dirac operator for a general
three-dimensional metric.  We also consider the compatibility of this
approach with ADM decomposition, which allows to extend this
continuous operator to the arbitrary four-dimensional metrics which
appear in GLET.  We have not yet found a nice lattice variant of this
operator for a general lattice.

The paper is organized as follows: First we consider continuous theory
for the Dirac operator on $(T\otimes\Omega)(\mathbb{R}^3)$, then the lattice
Dirac operator and the identification with standard model fermions.
Then we consider ether models.  Last not least we discuss gauge
fields.  Parts which seem essential for the understanding of the whole
picture but have been considered in earlier papers, especially GLET
\cite{SchmelzerAnnals}, Hodge theory and ADM decomposition
\cite{SchmelzerEP} and general remarks about quantization
\cite{SchmelzerGET}, we have shortly introduced in appendices.

\section{Continuous theory for the  Dirac operator 
on $(T\otimes\Omega)(\mathbb{R}^3)$}

Let's at first consider the lattice theory of the Dirac operator on
$(T\otimes\Omega)(\mathbb{R}^3)$ and its connection to standard model
fermions.  The Dirac operator on $\Omega(\mathbb{R}^3)$ is the Dirac
operator on the exterior bundle $\Omega(M^n)$ defined in Hodge theory
on arbitrary manifolds $M^n$ (we introduce this operator for this
general case in sec. \ref{Hodge}) for the case $M^n=\mathbb{R}^3$.

The Dirac operator used in QFT differs from the Hodge theory Dirac
operator in several questions: First, the Dirac operator in QFT has a
well-defined, fixed complex structure. For the Dirac operator on
$\Omega(M^n)$ no such structure is specified. We will see below
(sec. \ref{complex}) that several complex structures exist on
$\Omega(\mathbb{R}^3)$, which form a quaternionic structure.  This may
be used to define a natural complex structure on
$(T\otimes\Omega)(\mathbb{R}^3)$.  The complex structure is not important as
long as we consider the Dirac equation alone.  It becomes important
later, when we want to define interactions with gauge fields.

Then, the QFT Dirac operator has eight steps of freedom on
$(3+1)$-dimensional spacetime, while the Dirac operator on
$\Omega(M^n)$ has dimension $2^n$.  Thus, the necessary number of
steps of freedom for the QFT Dirac operator we obtain for
$\Omega(\mathbb{R}^3)$, not for
$\Omega(\mathbb{R}^3\otimes\mathbb{R})$.  Therefore the consideration
of the space $\Omega(\mathbb{R}^3)$ to describe fermions does not fit
into the relativistic spacetime paradigm.  It becomes much more
natural in a theory which handles space and time differently.  We
consider such proposals in sec. \ref{ether}.  In this context, we can
extend the three-dimensional operator using an essentially
three-and-one-half dimensional approach to a QFT Dirac operator.

\subsection{A Matrix representation of the Dirac operator 
on $\Omega(\mathbb{R}^3)$}

The matrix representation of the Dirac matrices $\gamma^\mu$ in
$\Omega(\mathbb{R}^3)$ may be defined in the following way:

\begin{equation}
\label{spinor}
(\gamma^0\partial_t-\gamma^i\partial_i) \varphi=_{def}
\left(
\begin{array}{rrrrrrrr}
 \partial_t& \partial_z& \partial_y&           & \partial_x&&&\\
-\partial_z&-\partial_t&           & \partial_y&& \partial_x&&\\
-\partial_y&           &-\partial_t&-\partial_z&&& \partial_x&\\
           &-\partial_y& \partial_z& \partial_t&&&& \partial_x\\           
-\partial_x&&&&-\partial_t&-\partial_z&-\partial_y&           \\
&-\partial_x&&& \partial_z& \partial_t&           &-\partial_y\\
&&-\partial_x&& \partial_y&           & \partial_t& \partial_z\\
&&&-\partial_x&           & \partial_y&-\partial_z&-\partial_t          
\end{array}
\right)
\left(
\begin{array}{r}
\varphi_{000}\\
\varphi_{001}\\
\varphi_{010}\\
\varphi_{011}\\
\varphi_{100}\\
\varphi_{101}\\
\varphi_{110}\\
\varphi_{111}
\end{array}
\right)
\end{equation}

where $\varphi = \sum \varphi_{ijk}(dx)^i(dy)^j(dz)^k$ is the
decomposition of an element $\varphi\in\Omega(\mathbb{R}^3)$.
In the context of this representation, it seems also natural to
define the following operators $\beta^i$ by their combination 
with coefficients $m_i$:

\begin{equation}
\label{betaidef}
m_i\beta^i =_{def}
\left(
\begin{array}{rrrrrrrr}
    & m_z& m_y&           & m_x&&&\\
 m_z&    &           & m_y&& m_x&&\\
 m_y&           &    &-m_z&&& m_x&\\
           & m_y&-m_z&    &&&& m_x\\           
 m_x&&&&    &-m_z&-m_y&           \\
& m_x&&&-m_z&    &           &-m_y\\
&& m_x&&-m_y&           &    & m_z\\
&&& m_x&           &-m_y& m_z&              
\end{array}
\right)
\end{equation}

The following operator equation holds:

\begin{equation}
(\gamma^0\partial_t-\gamma^i\partial_i + m_i\beta^i)^2 
 = -\square + \delta^{ij} m_i m_j
\end{equation}

This can be easily seen -- this operator iterates three times, in each
coordinate direction, the same trick:\footnote{This observation also
suggests how to iterate this construction to arbitrary dimension.}

\begin{equation}
\left(
\begin{array}{cc}
A&(m_i+\partial_i)I\\(m_i-\partial_i)I&-A
\end{array}
\right)^2
=
(A^2+(m_i+\partial_i)(m_i-\partial_i)I)\left(
\begin{array}{cc}
I&0\\0&I
\end{array}
\right)
\end{equation}

It follows, as it should, that the $\gamma^\mu$ define a
representation of the Dirac matrices.  It also follows immediately
that the matrices $\beta^i$ fulfill the following anti-commutation
relations:

\begin{equation}
\beta^i\beta^j+\beta^j\beta^i=\delta^{ij}
\end{equation}

and anti-commute with all $\gamma^\mu$:

\begin{equation}
\beta^i\gamma^\mu+\gamma^\mu\beta^i=0
\end{equation}

It is also easy to see that

\begin{equation}
\gamma^0(\gamma^1\beta^1)(\gamma^2\beta^2)(\gamma^3\beta^3)=1.
\end{equation}

\subsection{Complex structures on $\Omega(\mathbb{R}^3)$}
\label{complex}

This representation of the Dirac matrices does not have a complex
structure.  But such a structure seems necessary to define the
interaction with gauge fields, at least if we want to do it in the
standard way. 

In a real representation, a complex structure is defined by a linear
operator $i$.  The properties required for $i$ to define a complex
structure are are $i^{-1}=i^*=-i$ (where $i^*$ denotes the Euclidean
adjoint) and $[\gamma^\mu,i]=0$.  Now, an interesting point is that
there are several candidates for such a structure:

\begin{eqnarray}
i &= \beta^y\beta^z &= \iota\beta^x\\
j &= \beta^z\beta^x &= \iota\beta^y\\
k &= \beta^x\beta^y &= \iota\beta^z
\end{eqnarray}

which together define a quaternionic structure:\footnote{The classical
representation $ij=k$ can be obtained using reverse signs for $i,j,k$,
but we prefer this sign convention because it gives
$\gamma^5=\beta^x$.}

\begin{equation}
i j = - j i = -k;\; 
j k = - k j = -i;\; 
k i = - i k = -j;\; 
i^2 = j^2 = k^2 = -1
\end{equation}

We see that every complex structure is connected in a natural way with
a preferred direction.  This allows to define a natural complex
structure on $(T\otimes\Omega)(\mathbb{R}^3)$, where in each of the
three components $\Omega(\mathbb{R}^3)$ we use the preferred complex
structure of this direction.

This complex structure on $(T\otimes\Omega)(\mathbb{R}^3)$ no longer has a
preferred direction.  Nonetheless, it has a preferred orientation
(chirality).  

\subsection{Getting rid of unnecessary dependencies on the complex
structure}

As we see, the complex structure is a subtle issue.  As we will see
below, it becomes even more subtle on the lattice.  In this context,
it seems useful to clarify what really depends and what does not
depend on the choice of the complex structure, and to get rid of
unnecessary dependencies on the complex structure.

\subsubsection{The $\gamma^5$ operator}

The ``classical'' operator $\gamma^5=
-i\gamma^0\gamma^1\gamma^2\gamma^3$ depends on the complex structure.
A natural replacement which does not depend on it -- the expression
$\gamma^0\gamma^1\gamma^2\gamma^3$ -- we denote with $\iota$:

\begin{equation}
\iota=_{def}\gamma^0\gamma^1\gamma^2\gamma^3=\beta^x\beta^y\beta^z \hspace{1cm}
\iota\gamma^\mu+\gamma^\mu\iota=0\hspace{1cm}
\iota\beta^i = \beta^i\iota\hspace{1cm}
(\iota)^2 = -1
\end{equation}

For each candidate $i$ for a complex structure, we obtain an own
operator $\gamma^5=_{def}-i\iota$.  Especially for $i=\iota\beta^x$ we
obtain $\gamma^5=-\iota\beta^x\iota=\beta^x$. 

\subsubsection{Hermitian and Euclidean structure}

In the standard approach a Hermitian scalar product
$\langle.,.\rangle$ is widely used.  In our real representation we
have only a standard Euclidean scalar product $(.,.)$ yet.  Now, for a
complex structure $i$ these notions are closely related in a simple
way: The Hermitian scalar product defines an Euclidean scalar product
by

\begin{equation}
(\psi,\phi)=\frac{1}{2}(\langle\psi,\phi\rangle+\langle\phi,\psi\rangle
\end{equation}

so that $(i\psi,i\phi)=(\psi,\phi)$.  For a complex structure $i$,
$i^2=-1$, with this property this Hermitian scalar product is defined
by the Euclidean scalar product as

\begin{equation}
\langle\psi,\phi\rangle=(\psi,\phi)-i(\psi,i\phi).
\end{equation}

Thus, we should not care about the Hermitian scalar product, the
Euclidean scalar product is all we need.  

\subsubsection{Adjoint operators}

Another notion we can get rid of are Hermitian adjoint operators.  A
complex linear operator $A$ is simply a real linear operator with
$[A,i]=0$.  For such operators, the Hermitian adjoint operator $A^+$
and the Euclidean adjoint operator $A^*$ coinside: $\langle
A^*\psi,\phi\rangle=\langle\psi,A\phi\rangle$.  As a consequence, the
classical properties of the $\gamma$-matrices

\begin{equation}
(\gamma^\mu)^+ = \gamma^0\gamma^\mu\gamma^0
\end{equation}

are equivalent to

\begin{equation}
(\gamma^\mu)^* = \gamma^0\gamma^\mu\gamma^0.
\end{equation}

These properties are fulfilled in our representation for the standard
Euclidean scalar product $(.,.)$ in $\mathbb{R}^8$.

\section{Lattice theory of the Dirac operator on
$(T\otimes\Omega)(\mathbb{R}^3)$ and standard model fermions}

This representation is appropriate for a discretization of the Dirac
equation on a regular hyper-cubic lattice.  It can be obtained in a
quite simple way: We start with a naive central difference
approximation

\begin{equation}
\partial_i\varphi(n)\to\frac{1}{2h_i}(\varphi(n+h_i)-\varphi(n-h_i)).
\end{equation}

In \cite{SchmelzerNPB} we have seen that this naive discretization
gives, as it should be expected, $2^4=16$ fermions instead of one
(fermion doubling).  We have also found that these doublers decompose
in a simple way.  First, there is a decomposition into eight pairs:

\begin{equation}
\psi_{\kappa\lambda\mu} = (\varphi_{\kappa'\lambda'\mu'}\;\mbox{on
spatial node}\;[2k+\kappa+\kappa',2l+\lambda+\lambda',2m+\mu+\mu'])
\end{equation}

with $\kappa,\lambda,\mu,\kappa',\lambda',\mu' \in {0,1},
k,l,m\in\mathbb{Z}$.  Then, each pair decomposes into the part defined
on even and odd nodes of the spatial-temporal lattice.  Here, a node
$[k,l,m,t], k,l,m,t\in\mathbb{Z}$ is odd/even if $k+l+m+t$ is odd
resp. even.

In \cite{SchmelzerNPB} we have thrown away seven of the eight pairs,
leaving on the spatial node $(2k+\kappa,2l+\lambda,2m+\mu)$ only
$\varphi_{\kappa\lambda\mu}$.  On the other hand, we have suggested
not to throw away the last doubler.  Instead, we have proposed to
interpret the remaining pair of doublers as a physical flavor
doublet, as formed by the quarks of each color and generation or, if
the neutrino appears to be a Dirac particle, by leptons of each
generation.

In this paper, we consider the reverse approach: On one hand, we use
an ``accurate'' method, without doublers, for time.  For example, we
can do this in discrete time by hand, throwing away all odd nodes of
the space-time grid.  In the context of an ether theory (see
sec. \ref{ether}) this happens automatically: The natural discrete
structure (granular structure of polycrystalline ether, atomic
structure of atomic ether) gives only a spatial discretization but
leaves time continuous.  

On the other hand, we do not remove the remaining eight doublers and
suggest to interpret them as the eight fermions of a whole family (two
leptons, two quarks, each quark in three colors).

\subsection{Identification of the particles}

Now, the representation of the eight doublers in terms of
$\psi_{\kappa\lambda\mu}$ seems useful to understand some of their
symmetry properties.  In the free field limit all of them have
identical physical properties.  Indeed, they may be transformed into
each other by simple operators:

\begin{eqnarray}
\label{sidef}
s_x \psi_{\kappa\lambda\mu}(n) &=& \psi_{(1-\kappa)\lambda\mu}(n)\\
s_y \psi_{\kappa\lambda\mu}(n) &=& \psi_{\kappa(1-\lambda)\mu}(n)\\
s_z \psi_{\kappa\lambda\mu}(n) &=& \psi_{\kappa\lambda(1-\mu)}(n)
\end{eqnarray}

Let's consider now the eigenvectors of the commuting set of operators
$s_x,s_y,s_z$.  Because $s_x^2=s_y^2=s_z^2=1$, they have only
eigenvalues $\pm 1$.  The resulting eight eigenspaces have dimension
one and suggest a simple identification scheme with the fermions of
one standard model family\footnote{We use here symbols for the first
family only for the purpose of illustration.}:

\begin{eqnarray}
\nu &=&\sum_{\kappa\lambda\mu}\psi_{\kappa\lambda\mu}\\
d_r &=&\sum_{\kappa\lambda\mu}(-1)^\kappa  \psi_{\kappa\lambda\mu}\\
d_g &=&\sum_{\kappa\lambda\mu}(-1)^\lambda \psi_{\kappa\lambda\mu}\\
d_b &=&\sum_{\kappa\lambda\mu}(-1)^\mu     \psi_{\kappa\lambda\mu}\\
u_r &=&\sum_{\kappa\lambda\mu}(-1)^{\lambda+\mu}       \psi_{\kappa\lambda\mu}\\
u_g &=&\sum_{\kappa\lambda\mu}(-1)^{\mu+\kappa}        \psi_{\kappa\lambda\mu}\\
u_b &=&\sum_{\kappa\lambda\mu}(-1)^{\kappa+\lambda}    \psi_{\kappa\lambda\mu}\\
e   &=&\sum_{\kappa\lambda\mu}(-1)^{\kappa+\lambda+\mu}\psi_{\kappa\lambda\mu}
\end{eqnarray}

It seems remarkable that in this identification a color symmetry
appears almost automatically as rotational symmetry.  This
identification also gives a natural suggestion for the electromagnetic
charge of the particles: It may be identified with the number of
``sign factors'' of type $(-1)^\kappa$ in this definition.  Moreover,
there is a natural duality operation which connects the flavor pairs
connected by weak interactions.

\subsection{Some operators on fermions}

Let's define now some operators which seem to be useful for the
understanding of the lattice theory.  First, the following
operators seem to be interesting:

\begin{eqnarray}
\epsilon_x \psi_{\kappa\lambda\mu}&=&(-1)^\kappa  \psi_{\kappa\lambda\mu}\\
\epsilon_y \psi_{\kappa\lambda\mu}&=&(-1)^\lambda \psi_{\kappa\lambda\mu}\\
\epsilon_z \psi_{\kappa\lambda\mu}&=&(-1)^\mu     \psi_{\kappa\lambda\mu}
\end{eqnarray}

and $\epsilon = \epsilon_x\epsilon_y\epsilon_z$.  These operators
transform the fermions into each other.  Following the previous
particle identifications, the operators $\epsilon_i$ change
electromagnetic charge by $1/3$, while $\epsilon$ changes the flavor.

\subsection{Chiral symmetry on the lattice}

The problem with ``naive'' Dirac fermions as well as with standard
staggered fermions \cite{Kogut} is not only that they have the wrong
number of doublers (sixteen resp. four) to allow a natural physical
interpretation in the standard model.  The problem is also that there
is exact chiral $\gamma^5$ symmetry on the lattice.  As a consequence,
the doublers appear in pairs with reverse chiral charge.  This does
not fit the situation in the standard model (cf. \cite{Gupta98}).  Now,
in our approach we do not have exact chiral $\gamma^5$ symmetry.
Instead, we have a replacement for this symmetry.  This replacement
fulfills properties which define a generalization of the famous
Ginsparg-Wilson (GW) relation \cite{Ginsparg}.

Let's consider one fermion with family index x, thus, with complex
structure $i=\iota\beta^x$, $\gamma^5=\beta^x$.  To understand chiral
symmetry we have to define $\gamma^5=\beta^x$ on the lattice.  It
cannot be a pointwise operator as for Wilson fermions and staggered
fermions -- it connects components which are located in different
points.  Now, we propose to consider the following operator as a
candidate for $\gamma^5$ on the lattice:

\begin{eqnarray}
(\gamma^5 \phi)(n_{even}) &=& \phi(n_{even}-h_x)\\
(\gamma^5 \phi)(n_{odd }) &=& \phi(n_{odd }+h_x)
\end{eqnarray}

It is easy to see that it approximates the continuous $\gamma^5$.
More interesting is that some exact properties remain valid:

\begin{equation}
(\gamma^5)^*=\gamma^5; \;\;\;  (\gamma^5)^2=1
\end{equation}

We can also define, as an alternative, the operator $\tilde{\gamma}^5$
by

\begin{eqnarray}
(\tilde{\gamma}^5 \phi)(n_{even}) &=& \phi(n_{even}+h_x)\\
(\tilde{\gamma}^5 \phi)(n_{odd }) &=& \phi(n_{odd }-h_x)
\end{eqnarray}

Similarly, we obtain

\begin{equation}
(\tilde{\gamma}^5)^*=\tilde{\gamma}^5; \;\;\; (\tilde{\gamma}^5)^2=1
\end{equation}

If we define the operators $V,O$ by
$ \tilde{\gamma}^5 =  \gamma^5 V = \gamma^5(1-h_x O) $

we obtain the Ginsparg-Wilson (GW) relation for $O$:

\begin{equation}
O \gamma^5 + \gamma^5 O = h_x O \gamma^5 O
\end{equation}

Moreover, we have also the following important commutation properties
with $D$

\begin{eqnarray}
\tilde{\gamma}^5 D + D \gamma^5 &=& 0\\
\gamma^5 D + D \tilde{\gamma}^5 &=& 0\\
V D - D V &=& 0\\
O D - D O &=& 0
\end{eqnarray}

This allows to define two sets of chiral projector operators

\begin{eqnarray}
\tilde{P}_\pm &=& \frac{1}{2} (1\pm\tilde{\gamma}^5),\\
      {P}_\pm &=& \frac{1}{2} (1\pm      {\gamma}^5).
\end{eqnarray}

Similar pairs of projectors play a central role in approaches to
chiral gauge theory based on the GW relation (\cite{Golterman},
\cite{Luescher}) and it's generalizations (\cite{Kerler}) as domain
wall fermions \cite{Shamir}, Neuberger's overlap operator
\cite{Neuberger}, and proposals by Fujikawa \cite{Fujikawa} and Chiu
\cite{Chiu}.

On the other hand, there are some differences: The operators $V,
O$ do not have the spectral properties of the similar operators
considered, for example, by \cite{Golterman}, \cite{Kerler}.  At least
partially this difference may be understood as caused by different
aims.  The aim of the standard GW approach is to obtain a single Weyl
fermion on the lattice, without any doublers.  In our approach we do
not want to get rid of doublers at all.  Instead, we want nontrivial
chiral symmetry only to obtain nontrivial chiral interactions between
the doublers.

The more important difference is that in our approach the complex
structure is also not defined as a pointwise operator on the fermions.
This prevents the use of the standard Wilson approach to lattice gauge
theory and therefore also of the standard GW approach.  On the other
hand, we should not forget that this non-trivial character also
prevent the application of standard no-go theorems like the famous
Nielson-Ninomiya \cite{NN} theorem.

\section{Ether models for $(T\otimes\Omega)(\mathbb{R}^3)$ lattice theory}
\label{ether}

In the previous sections we have found a way to describe the fermionic
content of the standard model starting with the Dirac operator on
$(T\otimes\Omega)(\mathbb{R}^3)$ combined with ``naive'' spatial
discretization in space but not in time.  We find here several
occurrences of the number three -- the dimension of space, not of
space-time.  This suggests physical interpretation in terms of
theories which handle space and time differently.

We consider below two candidates for such theories: A polycrystalline
ether where the discrete structure of space is obtained by crystal
grains, and atomic ether theory where it is obtained by the atomic
structure.  In above cases, time is classical continuous absolute
time, without any discrete structure.  

\subsection{Polycrystalline ether theory}
\label{polyether}

In this section, we consider the derivation of
$(T\otimes\Omega)(\mathbb{R}^3)$ from a simple ``polycrystalline ether''
hypothesis.

The derivation itself is quite simple.  A polycrystalline material
consists of small crystallic grains.  If we want to do elasticity
theory for such a material, we have to describe distortions of such a
material.  If the grains are more rigid than the material between
them, the state of a grain $n$ may be described, in good
approximation, by the position of its center $q^i_0(n)$ and a linear
transformation $q^i_j(n)$.  Together with these state variables, we
also need related momentum variables $p^i_0(n)$, $p^i_j(n)$.  Now, we
can identify these steps of freedom with $(T\otimes\Omega)(\mathbb{R}^3)$ by
$q^i_0 + p^i_j dx^j + q^i_j *dx^j + p^i_0 dxdydz$.

This identification preserves not only the number of steps of freedom,
but preserves also important structural properties.  Duality between
configuration and momentum variables gives flavor doublets,
rotational symmetry gives color symmetry, three spatial directions
give three families.

\subsubsection{The special role of time}

An important point for the identification is that space and time
appear in the ether approach in a non-symmetric way: The
polycrystalline structure leads to a natural lattice structure in
space, but not in time.  Therefore, the lattice-related doubling
effect appears only in the three spatial directions.  Therefore we
obtain only $2^3=8$ doublers, instead of $2^4=16$ doublers as in a
naive space-time lattice.

\subsubsection{Empirical content of the polycrystalline ether hypothesis}

Starting from a single phrase -- ``polycrystalline ether'' -- we have
obtained not only the correct number of fermionic steps of freedom of
the standard model (192 real fields), but also the most important
structural properties of these steps of freedom ($192 = 3\cdot
2\cdot(1+3)\cdot 8$).  This identification does not look like
something made up.  The polycrystalline ether proposal seems very
restrictive at least in some parts.  For example, there would be no
possibility for a fourth fermion family, the neutrino should be a
standard Dirac particle.  This gives the theory sufficient empirical
content.  On the other hand, there will be large freedom in the choice
of various material parameters.  Therefore we should not expect
predictions of all SM parameters from this theory.

\subsubsection{Explanatory power}

The polycrystalline ether proposal is quite satisfactory also from
another point of view.  It is well understood how polycrystalline
materials may appear.  Especially they appear in a quite general
situation: Near second order phase transitions.  The concept of second
order phase transitions is a beautiful, attractive concept.  The
parameters which will be left unexplained seem to be almost as
unimportant from metaphysical point of view, comparable with the
parameters of planetary orbits in Newtonian gravity.  Whatever the
underlying microscopic theory, it will give some set of material
parameters which depends on the set of parameters of the underlying
microscopic theory.

\subsubsection{Open problems}

Of course, this state has not yet been reached.  What we have found
is, until now, only a partial success: the explanation of the
fermionic part of the standard model.

Already in the free field limit of dynamics -- the Dirac equation
without any gauge fields considered here -- we are faced with a
classical problem known already from classical ether theory: In usual
solid materials, including usual polycrystalline materials, we have
different speeds of sound for longitudinal and transversal waves.
This is a quite general property, it follows in standard elasticity
theory from quite simple symmetry considerations.  Instead, the Dirac
equation gives the same maximal speed for all types of excitations.

\subsection{Connection between Dirac and d'Alembert equation}
\label{Alembert}

Leaving this problem open, let's assume we have a polycrystalline
material which in some limit has the same d'Alembert equation equation
for all of its steps of freedom

\begin{equation}
\square q^i_0(n,t) = 0 \;\;\;\; \square q^i_j(n,t) = 0
\end{equation}

where $\square=\partial_t^2 - \Delta$ is the d'Alembert equation for
the discrete Laplace operator $\Delta$ on the lattice.  This second
order wave equation is already more close to equations for classical
polycrystalline materials than the first order Dirac equation.
Therefore it is worth to consider shortly their connection.

The three-dimensional lattice Dirac operator $D$ has the property
$D^2=\Delta$.  Each solution of the Dirac equation
$(\gamma^0\partial_t\pm D)\psi=0$ defines also a solution of the
d'Alembert equation.  In the other direction, the situation is less
trivial.  Assume we have a solution of the d'Alembert equation for
initial values $Q=\{q^i_0(n), q^i_j(n)\}$ and their first derivatives
$\dot{Q}=\{\dot{q}^i_0(n), \dot{q}^i_j(n)\}$.  This does not define in
general a unique solution $\{Q,P\}$ of the Dirac equation.  The
problematic part are homogeneous solutions $D\psi(n)=0$, especially
constants $\psi(n)=\psi_0=\{Q_0,P_0\}$.  This leads to non-uniqueness
because $P\to P+P_0$ does not change $\dot{Q}$, as well as
non-existence of $P$ for solutions of type $Q_0 t$.  But these
differences for constant solutions will not lead to physical effects
which are observable for internal observers.  Moreover we can suppress
them using appropriate boundary conditions.  In this sense, the
formulation using the Dirac equation can be considered as equivalent
to the formulation in terms of the d'Alembert equation.

Note that the shift from Dirac equation to d'Alembert equation may be
important for the understanding of symmetry breaking.  Indeed, the
Laplace operator $\Delta$ on the lattice has no preferred orientation.
Instead, the lattice Dirac operator depends on a choice of
orientation.  This choice of orientation is equivalent to the choice
of the sign of the square root in $D^2=\Delta$.

\subsection{Atomic ether theory}
\label{atomicether}

Most essential points of polycrystalline ether theory will be present
also in an alternative approach: Atomic ether theory.  An atomic ether
theory proposes some set of atoms which may be of different type.
They form a lattice.  Each type of atoms is described by a sub-lattice.
This gives, for $k$ types of atoms, $3k$ real steps of freedom.
Moreover, we have the same structure of space-time: The classical
equations are continuous in time but discrete in space.  Therefore we
obtain a doubling with factor $2^3=8$.

Comparison with the standard model gives now $k=4$.  Thus, we need
four types of atoms to describe the SM fermions.  This is a special
example of a general property: In comparison with the polycrystalline
theory, we have more freedom in the construction of the atomic theory.
This gives atomic ether theory less empirical content: It is much
easier to modify atomic theory to fit observation.  Nonetheless, it
may be that an atomic ether theory appears to be very simple,
comparable in simplicity with a polycrystalline ether theory.

The reason why we have introduced it here is that it allows to solve
the problem with the different speed of longitudinal and transversal
waves.  Indeed, if we have four atoms, we can assume strong forces
between atoms of the same type, and by Ockham's razor we would prefer
a theory where these forces do not depend on the type.  The free field
limit would be, in this case, the limit where we have no interaction
between atoms of different type, which leads to the same wave equation
for all four atoms.  Thus, in this atomic theory the d'Alembert
equation as a free field limit seems to be much more natural than in a
polycrystalline ether theory.

Let's mention another difference between these two variants of ether
theory.  There is a quite natural process in polycrystalline
materials: crystal grow.  This process changes the average distance
between the grains.  Now, such a change of the critical distance leads
to a renormalization which, from point of view of internal observers,
seems equivalent to an expansion of their universe.  Atomic ether
theory does not give such a natural mechanism for renormalization.

\section{Gauge Fields}

Gauge fields are not yet described in our ether proposal.  The main
reason for this is the unorthodox realization of the complex structure
on the lattice.  Indeed, while we have found a nice complex structure
in the continuous case, it does not define a pointwise complex
structure on the lattice.  Instead, the real and imaginary parts of a
``complex number'' of a fermion are located on different nodes.  Once
multiplication with $i$ is no longer a pointwise operation, the
standard Wilson approach to lattice gauge theory fails.

On the other hand, we have found some nice prerequisites which seem to
be useful to build such a theory.  We have already described operators
which seem to be related with electromagnetic charge, flavor and
color of the particles.  We have also found a nice nontrivial
realization of chiral symmetry on the lattice, similar to the lattice
version of chiral symmetry used in the Ginsparg-Wilson approach to
chiral gauge theory.

Moreover, there are some general principles about the nature of gauge
fields which follow from the ether approach: The gauge steps of
freedom should be handled like real steps of freedom.  Thus, there
should be some physical evolution equation for these steps of
freedom. The natural candidate is the Lorenz gauge.  Gribov copies
should be interpreted as really different field configurations.

\subsection{Some general principles for gauge fields in ether theory}

Let's consider here some general principles for the realization of
gauge fields in the context of an ether theory.  It appears that the
ether theory concept is already quite restrictive about these general
principles.

\subsubsection{Gauge degrees of freedom as physical steps of freedom}

There is the famous Bohm-Aharonov experiment which shows effects of
gauge fields in a region where the field strength $F_{\mu\nu}=0$.
This seems to exclude the possibility of description of gauge fields
using only the $F_{\mu\nu}$.  Thus, it seems necessary to use the
gauge potential $A_\mu$ to describe gauge fields.  Thus, we can assume
that the gauge potential $A_\mu$ or some equivalent on the lattice
(like integrals $\int A_\mu dx^\mu$ over edges) is used to describe
gauge steps of freedom.

Once non-gauge-invariant objects have to be used to describe steps of
freedom of the ether, they have to be physical steps of freedom.  This
has consequences:

\begin{itemize}

\item Quantization has to be done in the ``big'' space, without
factorization.

\item We have a definite Hilbert space structure on this space.

\item We need an evolution equation for the gauge degrees of
freedom.  Theories with different evolution equations are different as
ether theories, even if they appear to be indistinguishable by
observation. 

\item For this purpose, the Lorenz gauge condition is a natural
candidate, without any reasonable competitor.

\item Gribov copies are physically different states.

\end{itemize}  

\subsubsection{Definite Hilbert space structure}

In this context, it seems worth to note that the use of an indefinite
Hilbert space structure in the standard quantization approach has been
developed only 1950 by Gupta and Bleuer \cite{Gupta50}, \cite{Bleuer}
and is not the only possibility.  Instead, quantization is possible
also based on a standard Hilbert space.  In this way the
electromagnetic field has been initially quantized by Fermi
\cite{Fermi} and Dirac \cite{Dirac}.

The advantage of the Gupta-Bleuer approach is manifest Lorentz
symmetry.  But this advantage does not have much value in our
ether-theoretical approach, in comparison with a standard, physical
Hilbert space structure which gives manifest unitarity.  Indeed, if
something goes wrong with gauge symmetry, the use of an indefinite
Hilbert space leads to non-unitarity, and the resulting theory is
obviously physically meaningless.  If we, instead, start with a
definite Hilbert space, something may go wrong with Lorentz symmetry,
but the theory certainly remains to be physically meaningful.

Note that it is the ether approach which forces us to use such a
definite Hilbert space.  We have no choice here.  But the choice we
are forced to accept here seems to be a reasonable one.

\subsubsection{Lorenz gauge and Gribov copies}

The Lorenz gauge 

\begin{equation}
\partial_\mu A^\mu = 0 
\end{equation}

is not really a gauge, because it does not fix the gauge degrees of
freedom.  Instead, it is an evolution equation for them.  Indeed, for
a gauge transformation $A_\nu \to A_\nu + \partial_\nu \omega$ we
obtain

\begin{equation}
\partial_\mu g^{\mu\nu}\sqrt{-g} \partial_\nu \omega = 0 
\end{equation}

thus, the classical harmonic equation.  Thus, we have a whole field,
defined by arbitrary initial values $\omega(x,t_0)$,
$\partial_t\omega(x,t_0)$, which is not fixed by the Lorenz condition.

Now, this gauge degree of freedom should be interpreted as a physical
field.  Especially, Gribov copies define different physical states.

The choice of the Lorenz condition is also in good correspondence with
the approach used for gravity (see app. \ref{GLET}).  Indeed, in
gravity we use a similar coordinate condition -- the harmonic
condition -- which is interpreted as a physical equation.  Moreover,
the harmonic condition also has the form of a conservation law, and
this form has been used there to identify these equations with
classical conservation laws.  This analogy suggests not only the
choice of the Lorenz condition.  It also suggests to search for a
physical interpretation of the Lorenz condition in terms of
conservation of something.

\section{Summary}

What we have proposed here is a new paradigm for unification of all
fundamental forces of nature, based on completely different
metaphysics.  We revive the old ether idea in its full beauty: With
absolute space, absolute time, and an ether described by classical
condensed matter equations.  Relativistic symmetry as well as gauge
symmetry are not fundamental, but should be derived.  They are
secondary symmetries for observable effects, caused by the restricted
possibilities of internal observers, not true symmetries of reality
itself.

The ether paradigm essentially simplifies quantization (see
app. \ref{quantization}) and supports also a revival of classical
realism in quantum theory.  It is compatible with Bohmian mechanics
\cite{Bohm} and Nelson's stochastics \cite{Nelson} and defines a way
to extend them to theories of everything.

The general ether theory of gravity (app. \ref{GLET}) sufficiently
explains relativistic gravity, with interesting modifications: Frozen
stars instead of black holes, no big bang singularity, a dark matter
term.  The polycrystalline ether proposal made here derives the whole
fermionic content of the standard model, starting from almost nothing.
We obtain not only the correct number of degrees of freedom (192 real
fields), but all the basic structural properties: Eight components of
fermions, three families, flavor pairs, color symmetry between
quarks.  We have considered the free Dirac operator and found natural
operators which change flavor, color, electromagnetic charge and
chirality.  

A lot of interesting questions remain open.  The gauge sector has not
yet been understood.  We have found only some general principles for
handling gauge fields which differ from the standard relativistic
approach: Going back before Gupta-Bleuer, unitarity should be made
manifest, problems with unitarity in the relativistic approach should
be transformed into violations of Lorentz symmetry.  Gauge degrees of
freedom are physical, the Lorenz gauge is proposed as a physical
equation, Gribov copies understood as physically different
configurations. We have no Faddejev-Popov ghost fields.

Moreover, nothing has been done yet in the Higgs sector or for the
understanding of the fermion masses.  The modifications in the general
principles for gauge fields are too large to tell if or how the Higgs
mechanism has to be modified.

Despite these open problems, the initial success of the
polycrystalline ether hypothesis seems much too large to be
accidental.

\begin{appendix}

\section{Gravity}

A theory of everything should be able to describe gravity too.  We
consider here two questions: First, the way how to generalize a theory
based on the ``flat'' Dirac operator for $(T\otimes\Omega)(\mathbb{R}^3)$ to
a general four-dimensional metric background.  Second, we introduce a
general Lorentz ether theory which gives a metric theory of gravity
and seems compatible with the ether proposals made here for the
explanation of the standard model fermions.

The first question may be subdivided into two parts.  First, the
generalization of the three-dimensional Dirac operator on
$(T\otimes\Omega)(\mathbb{R}^3)$ to the case of a general three-dimensional
metric.  Here, the continuous theory is well-known standard Hodge
theory, we shortly remember the main results.  Unfortunately we have
not found yet a nice generalization of our ``naive'' lattice
discretization for a general lattice.  The second part is the use of
the ADM decomposition to extend the three-dimensional Dirac operator
to an essentially three-and-one-half-dimensional Dirac operator on a
four-dimensional space-time metric background.

The ether theory we propose can be understood as a modification of GR
which breaks covariance and fixes harmonic coordinates, combined with
an ether interpretation, which is based on the ADM decomposition and
the interpretation of the harmonic condition in terms of classical
conservation laws.  The theory identifies the observable gravitational
field with the classical energy-momentum tensor of the ether by
$g^{\mu\nu}\sqrt{-g}=t^{\mu\nu}$.  The material properties of the
ether, which are not specified in this theory, have to be identified
with the observable matter fields.

The most surprising observation is that this theory may be derived
from a few axioms which may be motivated from simple classical
principles.  Essentially, we need a Lagrange formalism and the
identification of the conservation laws given by Noether's theorem
with the conservation laws of classical condensed matter theory.  This
identification fixes four general equations, which are closely related
to the preferred coordinates.  They do not depend on the special
material properties of the ether.  As a consequence of the ``action
equals reaction'' symmetry of the Lagrange formalism, the equations
for these material properties of the ether do not depend on the
preferred coordinates.  But this is already the Einstein equivalence
principle.

\subsection{The Dirac operator on the de Rham complex}
\label{Hodge}

Until now we have used only a special case of the Dirac operator --
the Dirac operator on $\Omega(\mathbb{R}^3)$ with a standard Euclidean
metric.  We extend now this operator to a general metric background.
We need it only for a general Riemann metric on $\mathbb{R}^3$, but
this generalization is well-known from Hodge theory for a general
metric $g_{\mu\nu}(x)$ on a general manifold $M^n$ (see, for example,
\cite{Pete}).  Let's remember here the basic formulas:

The exterior bundle or de Rham complex $\Omega=\sum_{k=0}^n \Omega^k$
consists skew-symmetric tensor fields of type $(0,k), 0\le k \le n$
which are usually written as differential forms

\begin{equation}
\psi = \psi_{i_1\ldots i_k} dx^{i_1}\wedge \cdots \wedge dx^{i_k} \in \Omega^k
\end{equation}

The exterior bundle $\Omega$ has dimension $2^n$ in the
n-dimensional space.  The most important operation on $\Omega$ is
the external derivative $d:\Omega^k\to\Omega^{k+1}$ defined by

\begin{equation}
(d\psi)_{i_1\ldots i_{k+1}}=\sum_{q=1}^{k+1}\frac{\partial}{\partial x^{i_q}}
 (-1)^q \psi_{i_1\ldots \hat{i}_q\ldots i_{k+1}} 
\end{equation}

where $\hat{i}_q$ denotes that the index $i_q$ has been omitted. It's
main property is $d^2=0$.  In the presence of a metric, we have also
the important $*$-operator $\Omega^k\to\Omega^{n-k}$:

\begin{equation}
(*\psi)_{i_{k+1}\ldots i_n} = \frac{1}{k!} \varepsilon_{i_1\ldots i_n} 
g^{i_1j_1} \cdots g^{i_kj_k}\psi_{j_1\ldots j_k}
\end{equation}

with $*^2 = (-1)^{k(n-k)}\mbox{sgn}(g)$.  This allows to define a
global inner product by

\begin{equation}
(\phi,\psi) = \int \phi \wedge (*\psi) = \int \psi \wedge (*\phi)
\end{equation}

It turns out that the adjoint operator of $d^*$ of $d$ is

\begin{equation}
d^* = (-1)^{rn+n+1} * d *
\end{equation}

In this general context we can define the Laplace operator as

\begin{equation}
\Delta = d d^* + d^* d
\end{equation}

Then, the Dirac operator (as it's square root) can be defined as

\begin{equation}
D = d+d^*.  
\end{equation}

Indeed, we have $d^2=0$ as well as $(d^*)^2=0$.  

The ${\mathbb Z}_2$ graduation is also useful: $\varepsilon
\psi = (-1)^k\psi$ if $\psi\in\Omega^k$.  The subspaces
$\varepsilon=1$ and $\varepsilon=-1$ have equal dimension $2^{n-1}$.

The operator $\gamma^0$ in our representation can be understood as a
specialization of the graduation operator.  Indeed, the operator
$\varepsilon$ anti-commutes with the Dirac operator $D$.

Now, it would be nice to have a similar natural generalization of the
``naive'' discrete Dirac operator for a general lattice.
Unfortunately, the author has not found a nice generalization.  At the
current moment, the author favors the idea to get rid of the Dirac
equation in the regular situation, following sec. \ref{Alembert}.
Then what we have to generalize and to discretize on a general lattice
is only the d'Alembert equation, which is less problematic.

\subsection{Compatibility with ADM decomposition}

Our approach is in essential points three-dimensional.  Especially we
have used the Dirac operator on $\Omega(\mathbb{R}^3)$, combined with
an operator $\gamma^0$ defined by the graduation $\varepsilon$.

For the compatibility of the approach described here with a general
metric background we propose to use the ADM decomposition.  While this
consideration is independent of an ether interpretation, we
nonetheless use the denotations which we use later in our ether
theory.  Here they are simply nonstandard denotations for the standard
ADM decomposition:

\begin{eqnarray*} \label{ADM}
g^{00} \sqrt{-g} &=& \rho \\
g^{i0} \sqrt{-g} &=& \rho v^i \\
g^{ij} \sqrt{-g} &=& \rho v^i v^j + p^{ij}
\end{eqnarray*}

The ADM spatial coordinates are simply `comoving'' spatial coordinates
which remain constant along the ``velocity field''
$v^i=g^{0i}/g^{00}$.  The harmonic operator of the metric $g_{\mu\nu}$
in these coordinates reduces to

\begin{equation}
\square \psi = -(\rho \partial_t^2 - \Delta) \psi
\end{equation}

where $\Delta$ is already the standard three-dimensional (harmonic)
Laplace operator of the spatial metric.  This is already sufficient to
generalize the concept.  We can define now the four-dimensional Dirac
equation using the three-dimensional Dirac operator $D$ as

\begin{equation}
\sqrt{\rho} \varepsilon \partial_t \psi = \pm D \psi
\end{equation}

Moreover, in this decomposition we can also introduce a spatial
lattice which (following the comoving coordinates) moves continuously
in time.  Such a discretization gives, as required for our
identification, a doubling effect only in spatial directions.

\subsection{Definition of General Lorentz Ether Theory}
\label{GLET}

Let's introduce now an ether theory of gravity proposed in
\cite{SchmelzerProtvino} and described in more detail in
\cite{SchmelzerAnnals}, \cite{SchmelzerGET}.  
The theory has been named ``General Lorentz Ether Theory'' (GLET) for
three reasons: It generalizes the Lorentz ether to gravity, competes
with general relativity in a similar way as the Lorentz ether with
special relativity, and is also ``general'' as an ether theory because
it specifies only a few general properties of the ether, not its
material properties (which define the matter content).

GLET preserves the essential features of classical ether theories: We
have a classical Newtonian framework of absolute Euclidean space and
absolute time.  The space is filled with an ether.  This ether is
described using some general condensed matter variables: positive
density $\rho(x)$, velocity $v^i(x)$, and a pressure tensor
$p^{ij}(x)$, as well as some other material properties $\varphi^m(x)$.
These other properties are not specified by this general theory.  They
have to be specified in special ether models.  The pressure tensor is
proposed to be negative definite.  The gravitational field is defined
by the classical energy-momentum tensor of the ether in a variant of
ADM decomposition:

\begin{eqnarray*} \label{gdef}
g^{00} \sqrt{-g} &= t^{00} =& \rho \\
g^{i0} \sqrt{-g} &= t^{i0} =& \rho v^i \\
g^{ij} \sqrt{-g} &= t^{ij} =& \rho v^i v^j + p^{ij}
\end{eqnarray*}

The equations of the theory are Euler-Lagrange equations for the
Lagrangian

\begin{equation}
L = L_{GR}(g_{\mu\nu}) + L_{matter}(g_{\mu\nu},\varphi^m)
  - (8\pi G)^{-1}(\Upsilon g^{00}-\Xi (g^{11}+g^{22}+g^{33}))\sqrt{-g}
\end{equation}

where $L_{GR}$, $L_{matter}$ are the most general covariant
Lagrangians known from classical general relativity.  The additional
background-dependent terms break general covariance and fix a
coordinate condition. The preferred Newtonian coordinates $X^\mu$ are
the harmonic coordinates:

\begin{equation} \square X^\nu = \partial_\mu (g^{\mu\nu}\sqrt{-g}) = 0\end{equation}

Rewritten in the original ether variables, these conditions become the
classical conservation laws -- continuity equation

\begin{equation} \label{continuity}
 \partial_t \rho + \partial_i (\rho v^i) = 0
\end{equation}

and Euler equation

\begin{equation} \label{Euler}
 \partial_t (\rho v^j) + \partial_i(\rho v^i v^j + p^{ij})=0
\end{equation}

From this point of view, the theory looks like a minor modification of
GR.  An additional term breaks covariance and fixes harmonic
coordinates.  This is combined with an ether interpretation which is
surprisingly close to classical condensed matter theory.

\subsection{Axioms of General Lorentz Ether Theory}

Much more interesting is that the Lagrangian of the theory, and
especially the covariance of the matter Lagrangian (that means, the
Einstein equivalence principle) may be derived from simple ether
theory axioms.  We will shortly introduce them, for a more detailed
consideration see \cite{SchmelzerAnnals}.

The formulation of the main axiom as well as the derivation is based
on an unorthodox variant of the Lagrange formalism and Noether's
theorem as well as a slightly unorthodox choice of variables for the
ether.

\begin{itemize}

\item We have a subdivision into the energy-momentum variables 
$t^{\mu\nu}$ and other material steps of freedom $\varphi^m$, with the
$t^{\mu\nu}$ as independent variables.  Instead, in classical
condensed matter theory, pressure $p^{ij}$ is usually defined as a
given function of the other material steps of freedom, therefore, the
$t^{\mu\nu}$ are not independent from the $\varphi^m$.

\item We require that dependence on the preferred coordinates of the
Newtonian background $X^\mu$ is made explicit.

\end{itemize}

This point needs explanation.  The preferred coordinates $X^\mu$ are
also, as all physical fields, functions $X^\mu(x)$ on space-time.
Expressions can depend on them in explicit and implicit ways.  An
example is the expression $u^\mu T_{,\mu}$, which explicitly depends
on $T$.  It is equivalent to $u^0$, which therefore also depends on
$T$ -- but this dependence is no longer explicit.

In general, we name the coordinate-dependence of an expression
$F(\phi,\phi_{,\mu},\ldots,X^\alpha,X^\alpha_{,\mu},\ldots)$ explicit
if after replacement of the occurrences of $X^\alpha(x)$ by four scalar
fields $U^\alpha(x)$ no coordinate-dependence is left.  Thus, the
expression
$F(\phi,\phi_{,\mu},\ldots,U^\alpha,U^\alpha_{,\mu},\ldots)$ should be
covariant.

Once the coordinate-dependence in the Lagrangian is explicit, we can
vary over the coordinates $X^\mu$ and obtain Euler-Lagrange equations
for them by the standard rules.\footnote{Note that only variations
$\delta X^\mu$ so that $X^\mu+\delta X^\mu$ define valid coordinates
are allowed.  But for any $\delta X^\mu(x)$ with finite support this
holds for $\varepsilon\delta X^\mu(x)$ for small enough $\varepsilon$.
Therefore this does not restrict the variation.}  

If we have translational symmetry $X^\mu(x)\to X^\mu+c^\mu$ of the
Lagrangian, Noether's theorem tells that we obtain conservation laws.
But in our formalism we don't need Noether's theorem to find them.
Instead, the Euler-Lagrange equations for $X^\mu$ are conservation
laws: Indeed, $\partial L/\partial X^\mu = 0$, and all other terms
appear in the Euler-Lagrange equations under some partial
derivative.\footnote{In this formalism it is also easy to understand
why these conservation laws disappear if the Lagrangian is covariant.
In this case, the Euler-Lagrange equation for the $X^\mu$ disappears.}

Now we can formulate our main axiom: We require that the
Euler-Lagrange equation for the preferred coordinates are proportional
to the classical energy-momentum conservation laws.  In the preferred
coordinates we obtain:

\begin{equation}
\label{mainaxiom}
\frac{\delta S}{\delta X^\mu} = \gamma_{\mu\nu} \partial_\kappa
t^{\kappa\nu}
\end{equation}

for some constant matrix $\gamma_{\mu\nu} = 4\pi
G\mbox{diag}(-\Upsilon,\Xi,\Xi,\Xi)$.

\subsection{Derivation of the Lagrangian}

The expression on the right side of equation \ref{mainaxiom}
contains an implicit dependence on the coordinates -- the index $\nu$
in $t^{\kappa\nu}$.  Making it explicit, and using the metric
variables $g^{\mu\nu}\sqrt{-g}=t^{\mu\nu}$ gives

\begin{equation}
\frac{\delta S}{\delta X^\mu} = \gamma_{\mu\nu} \partial_\kappa
(g^{\kappa\lambda}\sqrt{-g} X^\nu_{,\lambda}) \equiv \gamma_{\mu\nu} \square X^\nu
\end{equation}

for the harmonic operator $\square$ of the metric $g^{\mu\nu}$.  There
is a simple particular solution for this:

\begin{equation} 
L_{part}= \gamma_{\mu\nu} g^{\kappa\lambda}\sqrt{-g}X^\mu_{,\kappa}
X^\nu_{,\lambda}
\end{equation}

For the general solution follows

\begin{equation}
\frac{\delta S-S_{part}}{\delta X^\mu}= 0,
\end{equation}

which defines the Lagrangian of general relativity in its most general
form:

\begin{equation}
L-L_{part} = L_{GR}(g_{\mu\nu}) + L_{matter}(g_{\mu\nu},\varphi^m)
\end{equation}

Especially, because $L_{part}$ does not depend on the variables
$\varphi^m$, the whole ``matter Lagrangian'' should be covariant.

\subsection{Physical predictions}  

Some differences between this ether theory and general relativity are
considered in \cite{SchmelzerAnnals}.  The most interesting one seems
that for $\Upsilon>0$ the gravitational collapse stops immediately
before horizon formation.  Thus, there are no black holes, but frozen
stars.  For sufficiently small $\Upsilon>0$ this does not seem to lead
to really observable differences.  But the differences are at least
theoretically important.  Especially there will be no Hawking
radiation.

Ether theory predicts a flat universe, in good agreement with
observation.  The $\Xi$-term gives an interesting cosmological term
for dark energy.  $\Upsilon>0$ prevents the big bang singularity.  The
solution for a flat Robertson-Walker universe will be a ``big
bounce'', but it remains open if this big bounce survives if we take
into account inhomogeneities and viscosity.  For details, see
\cite{SchmelzerGET}, \cite{SchmelzerAnnals}.

\section{Quantization}\label{quantization}

The quantization of gravity is considered to be one of the greatest
problems of modern science.  From point of view of ether theory, it
seems hard to understand what is problematic here.  Most hard problems
disappear. If an atomic ether theory, which defines a natural
regularization, is given, quantization itself seems possible and quite
unproblematic following the classical canonical quantization scheme.
Thus, the problem seems to be more a problem of definition of an
appropriate classical atomic ether theory which gives the classical
continuous ether theory in the large distance limit than a
quantization problem.

\subsection{Relativistic quantization problems which disappear in
ether theory}

Quantization of an ether theory in a classical Newtonian framework is
certainly a much simpler job than quantization of general relativity.
The list of quantization problems of GR quantization which disappear
in ether theory includes:

\begin{itemize}

\item The problem of time \cite{Isham}:  Once we have absolute time,
we have no such problem.

\item The information loss problem \cite{Preskill}: We have no black
holes but stable frozen stars, without Hawking radiation.

\item Quantum uncertainty of the light-cone and, therefore, of
relativistic causality: Causality in ether theory is classical
causality, connected with absolute time, and does not become
uncertain.

\item Closed causal loops: They do not appear in ether theory because
they violate the condition $\rho>0$.  If $\rho\to 0$ the field theory
limit fails, but not the fundamental atomic ether theory which is
quantized.

\item Physical meaning of the Hamiltonian constraint: We have a 
classical Hamilton formalism because we have classical absolute time.

\end{itemize}  

\subsection{Quantization of first order equations}

In quantum field theory, the quantization of first order equations
like the Lorenz gauge condition or the harmonic coordinate condition
is considered to be especially problematic.  Because we propose to use
these conditions in our ether theory, let's consider how they may be
quantized.

Assume we have a classical atomic model of the ether.  For the
quantization of this model we do not have to use any field theory,
multi-particle Schr\"odinger theory is sufficient.  If we have defined
this multi-particle theory, we have obtained a well-defined quantum
theory of this ether.  Now, conservation of particle number leads, in
the classical large distance limit, to a continuity equation:

\begin{equation}
\partial_t \rho + \partial_i \rho v^i = 0 
\end{equation}

Note, especially, that this conservation law holds exactly, there are
no ``quantum fluctuations'' of the particle number in multi-particle
Schr\"odinger theory.  Thus, quantization of such first order
equations is not only possible, but a standard feature of atomic
theories.  What we need to quantize such an equation is, therefore, an
atomic model so that the equation in question is interpreted as the
conservation of some number of atoms.

This does not mean that the situation in the quantum field theory
limit is nice.  It is not.  Nontrivial problems appear also in quantum
field theory for usual condensed matter.  For example, the definition
of the operator $\hat{\rho}$ is not unproblematic.  Especially, the
canonical commutation relations proposed by Landau \cite{Landau}

\begin{equation}
[\hat{\bf u}(x,t), \hat{\rho}(x',t)] = -i\hbar \nabla \delta(x-x')
\end{equation}

lead to a continuous, unbounded spectrum for $\hat{\rho}$ which is
incompatible with it's positivity \cite{Wagner}.  Such observations
suggests that there may be a lot of problems of the field theory
limit.  But these are not problems of the fundamental atomic ether
theory, nor are they quantization problems.

\subsection{EPR realism and hidden variables}

If we quantize some atomic ether theory using classical multi-particle
Schr\"odinger theory, we can easily extend this to obtain hidden
variable theories like Bohmian mechanics or Nelson's stochastics which
have been defined in this classical framework.  Thus, ether theory, if
successful, allows to generalize these hidden variable theories to a
theory of everything.

On the other hand, these hidden variable theories, in combination with
Bell's theorem and the observed violation of Bell's inequality, give
independent strong support for the existence of a preferred frame.
Indeed, in Bell's theorem Einstein causality and EPR realism are the
only ingredients we need.  It's violation proves that one of these two
principles is wrong. Thus, we have a conflict between Einstein
causality and EPR-realism.

In this conflict, EPR-realism is the more fundamental principle: It is
easy to imagine a world without Einstein causality, but a world which
is not EPR-realistic contradict common sense.  While we agree that
disagreement with common sense is not a decisive argument, it is
certainly a very strong argument and cannot be simply dismissed.

Moreover, the existence of EPR-realistic hidden variable theories for
quantum theory proves that EPR-realism is compatible with quantum
theory.  Therefore, quantum theory cannot give an independent argument
for the rejection of EPR-realism.  On the other hand, the existing
problems of GR quantization give such independent evidence against
Einstein causality.  Thus, the question ``which principle is
compatible with quantum principles'' clearly favors EPR-realism.

Therefore, it seems much more reasonable to reject Einstein causality
than EPR-realism.

The situation can be reformulated in another way: If we start with
EPR-realism as an axiom, Einstein causality is falsified, and we can
derive that there exists a preferred frame.  These considerations can
be found, in more detail, in \cite{SchmelzerGET}.

\end{appendix}

\end{document}